\documentclass[article,nofootinbib,superscriptaddress,notitlepage]{revtex4-1}
\usepackage{graphicx}
\usepackage{slashed}
\usepackage{amsmath}
\usepackage{enumerate}
\def\beq{\begin{equation}}
\def\eeq{\end{equation}}
\def\bea{\begin{eqnarray}}
\def\eea{\end{eqnarray}}

\def\tev{\text{ TeV}}
\def\gev{\text{ GeV}}

\def\fb{\text{ fb}}
\def\invfb{\text{ fb}^{-1}}

\def\sigmaSI{\sigma_{\rm SI}}
\def\sigmaSD{\sigma_{\rm SD}}

\begin{document}

\preprint{UH511-1247-2015}
\preprint{CETUP2015-017}

\title{Vector dark matter at the LHC}

\author{Jason Kumar}
\affiliation{Department of Physics \& Astronomy, University of Hawai'i, Honolulu, HI 96822, USA}

\author{Danny Marfatia}
\affiliation{Department of Physics \& Astronomy, University of Hawai'i, Honolulu, HI 96822, USA}

\author{David Yaylali}
\affiliation{Department of Physics, University of Arizona, Tucson, AZ 85721, USA}
\affiliation{Department of Physics, University of Maryland, College Park, MD 20742, USA}



\begin{abstract}
We consider monojet searches at the Large Hadron Collider (LHC) for spin-1 dark matter that interacts with quarks through a contact operator.  If the dark matter particles are produced with longitudinal polarizations, then the production matrix element is enhanced by factors of the energy. We show that this particularly effective search strategy can test models for which the energy suppression scale of the operator is as large as $10^5 \tev$.  As such, these searches can probe a large class of models for which the contact-operator approximation is valid.  We find that for contact operators that permit velocity-independent dark matter--nucleon scattering, LHC monojet searches for spin-1 dark matter are competitive with or far surpass direct-detection searches depending on whether the scattering is spin independent or spin dependent, respectively.
\end{abstract}

\pacs{95.35.+d}

\maketitle


\section{Introduction}

Broadly speaking, there are two search strategies for dark matter (DM) at hadron
colliders: cascade decay searches and direct-production searches.  The goal of
cascade decay searches is to produce a heavy QCD-charged particle whose decays
are required by symmetry to produce dark matter, and the prototypical examples are
supersymmetry searches for squark or gluino production (see, for example,~\cite{Aad:2015pfx,Aad:2015iea,Khachatryan:2015lwa}).  The goal of direct production
searches is to produce dark matter through a four-point (not necessarily contact)
interaction that couples two dark matter particles directly to partons.  The prototypical
examples of this strategy are  ``mono-anything" searches~\cite{Feng:2005gj,Goodman:2010ku,Bai:2010hh,Goodman:2010yf,Cheung:2012gi,Bell:2012rg,Zhou:2013fla}
in which one searches for a
single jet, photon, $W$, or $Z$ radiated from the partons, against which the DM
particles can recoil.

Cascade decay searches are most sensitive to the properties of the heavy
particles that are initially produced, whereas it is typically more difficult to
indirectly determine the properties of the DM decay product.  On the other
hand, although direct searches may have the disadvantage of not utilizing a strongly
coupled production process, they can more directly probe the properties of the DM particle and its interactions with the Standard Model (SM).
Direct searches have been used extensively at the LHC for the case where dark matter
is a spin-0 or spin-1/2 particle~\cite{ATLAS:2012ky,Chatrchyan:2012me,ATLAS,Aad:2013oja,CMS,Aad:2014vka}.
The goal of this work is to discuss the unique
signatures arising from the case of spin-1 dark matter (which we denote
by $B$).

A characteristic feature of vector dark matter is that the longitudinal polarization
vector scales as $\sim E / m$, implying that the production matrix element can receive a
significant enhancement in the region of phase space where the dark matter is relativistic and
either one or both particles are longitudinally polarized~\cite{Kumar:2013iva}.  The large enhancement of the
matrix element results in large cross sections for the process $pp \rightarrow B^\dagger B + \text{jet}$, yielding
greater sensitivity to the dark matter-quark coupling.

The plan of this paper is as follows.  In Section II, we tabulate the set of contact operators
that we use to model dark matter--quark interactions, and determine the energy scaling of the
DM production cross section that would be expected from considerations of angular momentum,
charge conjugation, and parity.
In Section III we discuss the limits on the applicability of the contact-operator approximation that
arise from unitarity of the scattering amplitude.  In Section IV we present the exclusion bounds on these models
imposed by results of the 8 TeV LHC run, the future sensitivity that may be expected from the
14 TeV LHC run, and the implications for dark matter--nucleon scattering.  We conclude with a discussion of our results in Section V.

\section{Operators and Matrix Elements}
\label{sec:Operators}

For simplicity we work under the assumption that spin-1 dark matter interacts with quarks
through effective contact operators.
The most general set of contact operators that we can write, up to dimension 6, is given in Table~\ref{Table:Operators}.
Note that we have assumed that $B$ is a complex vector field.
If $B$ is a real field ($B=B^\dagger$), then operators V$(3-6)$ and V$(7-10)_-$ vanish identically.
The remaining operators must be scaled by a factor of $1/2$ if $B$ is real.
We follow notation similar to~\cite{Kumar:2013iva}, wherein the operators V$(7-10)$ are herein
referred to as V$(7-10)_+$.\footnote{Note, Ref.~\cite{Kumar:2013iva} did not consider the operators V$(7-10)_-$.  For
completeness, we redo much of the analysis of Ref.~\cite{Kumar:2013iva} for these operators in Appendix C.}

\begin{table}[t]
  \begin{tabular}{ l c c }
    \hline \hline
    Operator & Structure & Dim $D$ \\ \hline
    V$1$ &  $(1 / \Lambda) B_{\mu}^{\dagger} B^{\mu} \bar{q} q$  & 5 \\ [5pt]
    V$2$ &  $(1 / \Lambda) \imath B_{\mu}^{\dagger} B^{\mu} \bar{q} \gamma^5 q$ & 5 \\ [5pt]
    V$3$ & $(1 / 2 \Lambda^2) \imath (B^{\dagger}_{\nu} \partial_{\mu} B^{\nu} - B^{\nu} \partial_{\mu} B^{\dagger}_{\nu}) \bar{q} \gamma^{\mu} q$ & 6 \\ [5pt]
    V$4$ & $(1 / 2 \Lambda^2) \imath (B^{\dagger}_{\nu} \partial_{\mu} B^{\nu} - B^{\nu} \partial_{\mu} B^{\dagger}_{\nu}) \bar{q} \gamma^{\mu} \gamma^5 q$ & 6  \\ [5pt]
    V$5$ & $(1 / \Lambda) \imath B_{\mu}^{\dagger} B_{\nu}  \bar{q} \sigma^{\mu \nu} q$ & 5  \\ [5pt]
    V$6$ & $(1 / \Lambda) B_{\mu}^{\dagger} B_{\nu}     \bar{q} \sigma^{\mu \nu} \gamma^5 q$ & 5   \\ [5pt]
    V$7_+$ & $(1 / 2 \Lambda^2) (B_{\nu}^{\dagger} \partial^{\nu} B_{\mu} + B_{\nu} \partial^{\nu} B_{\mu}^{\dagger}  )\bar{q} \gamma^{\mu} q$  & 6  \\
    V$7_-$ & $(1 / 2 \Lambda^2) \imath (B_{\nu}^{\dagger} \partial^{\nu} B_{\mu} - B_{\nu} \partial^{\nu} B_{\mu}^{\dagger}  )\bar{q} \gamma^{\mu} q$  & 6  \\[5pt]
    V$8_+$ & $(1 / 2 \Lambda^2) (B_{\nu}^{\dagger} \partial^{\nu} B_{\mu} + B_{\nu} \partial^{\nu} B_{\mu}^{\dagger}) \bar{q} \gamma^{\mu} \gamma^5 q$ & 6  \\
    V$8_-$ & $(1 / 2 \Lambda^2) \imath (B_{\nu}^{\dagger} \partial^{\nu} B_{\mu} - B_{\nu} \partial^{\nu} B_{\mu}^{\dagger}) \bar{q} \gamma^{\mu} \gamma^5 q$ & 6  \\ [5pt]
    V$9_+$ & $(1 / 2 \Lambda^2) \epsilon^{\mu \nu \rho \sigma} (B_{\nu}^{\dagger} \partial_{\rho} B_{\sigma} + B_{\nu} \partial_{\rho} B_{\sigma}^{\dagger})
    \bar{q} \gamma_{\mu} q$  & 6  \\
    V$9_-$ & $(1 / 2 \Lambda^2) \imath \epsilon^{\mu \nu \rho \sigma} (B_{\nu}^{\dagger} \partial_{\rho} B_{\sigma} - B_{\nu} \partial_{\rho} B_{\sigma}^{\dagger})
    \bar{q} \gamma_{\mu} q$  & 6  \\ [5pt]
    V$10_+$ & $(1 / 2 \Lambda^2) \epsilon^{\mu \nu \rho \sigma} (B_{\nu}^{\dagger} \partial_{\rho} B_{\sigma} + B_{\nu} \partial_{\rho} B_{\sigma}^{\dagger})
    \bar{q} \gamma_{\mu} \gamma^5 q$ & 6  \\
    V$10_-$ & $(1 / 2 \Lambda^2) \imath \epsilon^{\mu \nu \rho \sigma} (B_{\nu}^{\dagger} \partial_{\rho} B_{\sigma} - B_{\nu} \partial_{\rho} B_{\sigma}^{\dagger})
    \bar{q} \gamma_{\mu} \gamma^5 q$ & 6  \\
    \hline \hline
  \end{tabular}
  \caption{Possible Hermitian contact operators up to dimension 6 that couple spin-1 dark matter to SM quarks (or other fermions).}
\label{Table:Operators}
\end{table}

The operators in Table~\ref{Table:Operators} are assumed to be the low energy manifestations of some more fundamental ultraviolet theory. 
All of the contact operators given above can arise from renormalizable interactions in which the dark matter pair is produced 
by the $s$-channel exchange of a spin-1 or spin-0 mediator (for example, a heavy $Z'$ or Higgs particle), or by the 
$t$-/$u$-channel exchange of a spin-1/2 particle.  A detailed analysis of these constructions is presented in~\cite{Agrawal:2010fh}.
Our intention is to perform a completely general analysis of spin-1 dark matter at the LHC;  
under the assumption that the effective theory description is valid at these energies, the operators in Table~\ref{Table:Operators} 
represent a basis set to describe these interactions.

We are interested in the scaling of the matrix element for the process $\bar q q \rightarrow B^\dagger B$ with
respect to the energy $E$ in the center-of-mass frame of the $\bar q q$ system.  This scaling is determined by the
following considerations:
\begin{itemize}
\item{The matrix element scales by a factor $(E/\Lambda)^{d-4}$, where $d$ is the dimension of the
operator and $\Lambda$ is the energy scale of the coefficient. }
\item{The matrix element also scales by additional factors of $E/m_B$ for each DM
longitudinal polarization vector.  The number of such vectors can be found by determining the $C$, $P$,
and $J$ quantum numbers of the DM state that can be created by each operator.  These in turn
determine the $L$ and $S$ quantum numbers of the DM final state, which determine the polarization vectors.}
\end{itemize}

The $C$ and $P$ quantum numbers of the dark matter two-particle state (for this purpose, the jet(s) in the
final state are irrelevant) created by each operator are determined by the transformation properties of the DM
bilinear.  The $J$ quantum number of the dark matter two-particle state is determined by the rotational transformation
properties of the DM bilinear.  The $L$ and $S$ quantum numbers of the
DM final state system are then given by
\begin{eqnarray}
C: (-1)^{L+S}, \qquad P: (-1)^L , \qquad |L-S| \leq J \leq |L+S| ,
\end{eqnarray}
which are valid when the dark matter is a boson.
Following~\cite{Kumar:2013iva}, in Table~\ref{Table:CPJLS} we display the quantum numbers of the DM state created by the various terms in the above operators.
Note that we have ignored all operator terms involving the quark bilinears $\bar q \gamma^0 q$ and
$\bar q \gamma^0 \gamma^5 q$.  The former vanishes identically when acting on any quark-antiquark
initial state, while the latter yields a matrix element that is proportional to $m_q$, and thus vanishes
in the limit $m_q \ll E$, as is relevant here.

\begin{table}[h!]
  \begin{tabular}{ l  l  c  c  c  c }
    \hline \hline
    Operator & Term & $C_B$ & $P_B$ & \, J \, & State  \\
    \hline
    V$1$ &  $(1 / \Lambda) B_{\mu}^{\dagger} B^{\mu} \bar{q} q$  &  + & + & 0 & $L=0$, $S=0$; $L=2$, $S=2$  \\  [5pt]
    V$2$ &  $(1 / \Lambda) \imath B_{\mu}^{\dagger} B^{\mu} \bar{q} \gamma^5 q$ & + & + & 0 & $L=0$, $S=0$; $L=2$, $S=2$  \\ [5pt]
    V$3$ & $(1 / 2 \Lambda^2) \imath (B^{\dagger}_{\nu} \partial_{i} B^{\nu} - B^{\nu} \partial_{i} B^{\dagger}_{\nu}) \bar{q} \gamma^{i} q$  & - & - & 1 & $L=1$, $S=0$; $L=1,3$, $S=2$ \\ [5pt]
    V$4$ & $(1 / 2 \Lambda^2) \imath (B^{\dagger}_{\nu} \partial_{i} B^{\nu} - B^{\nu} \partial_{i} B^{\dagger}_{\nu}) \bar{q} \gamma^{i} \gamma^5 q$ & - & - & 1 & $L=1$, $S=0$; $L=1,3$, $S=2$   \\ [5pt]
    V$5$ & $(1 / \Lambda) \imath B_i^{\dagger} B_j  \bar{q} \sigma^{ij} q$ & - & + & 1 & $L=0,2$, $S=1$  \\
     & $(1 / 2\Lambda) \imath (B_0^{\dagger} B_i - B_i^{\dagger} B_0  \bar{q})  \bar{q} \sigma^{0i} q$ & - & - & 1 & $L=1$, $S=0$; $L=1,3$, $S=2$   \\ [5pt]
    V$6$ & $(1 / \Lambda) B_i^{\dagger} B_j     \bar{q} \sigma^{ij} \gamma^5 q$ & - & + & 1 & $L=0,2$, $S=1$    \\
    & $(1 / 2\Lambda) (B_0^{\dagger} B_i - B_i^{\dagger} B_0  )   \bar{q} \sigma^{0i} \gamma^5 q$ & - & - & 1 & $L=1$, $S=0$; $L=1,3$, $S=2$   \\ [5pt]
    V$7_+$ & $(1 / 2\Lambda^2) (B_{\nu}^{\dagger} \partial^{\nu} B_i + B_{\nu} \partial^{\nu} B_i^{\dagger}  )\bar{q} \gamma^i q$   &
    + & - & 1 & $L=1$, $S=1$   \\ [5pt]
    V$7_-$ & $(1 / 2\Lambda^2) \imath (B_{\nu}^{\dagger} \partial^{\nu} B_i - B_{\nu} \partial^{\nu} B_i^{\dagger}  )\bar{q} \gamma^i q$   &
    - & - & 1 & $L=1$, $S=0$; $L=1,3$, $S=2$  \\ [5pt]
    V$8_+$ & $(1 / 2\Lambda^2) (B_{\nu}^{\dagger} \partial^{\nu} B_i + B_{\nu} \partial^{\nu} B_i^{\dagger}) \bar{q} \gamma^i \gamma^5 q$ &
    + & - & 1 & $L=1$, $S=1$  \\ [5pt]
    V$8_-$ & $(1 / 2\Lambda^2) \imath (B_{\nu}^{\dagger} \partial^{\nu} B_i - B_{\nu} \partial^{\nu} B_i^{\dagger}) \bar{q} \gamma^i \gamma^5 q$ &
    - & - & 1 & $L=1$, $S=0$; $L=1,3$, $S=2$  \\ [5pt]
    V$9_+$ & $(1 / 2\Lambda^2) \epsilon^{i 0jk} (B_{0}^{\dagger} \partial_{j} B_{k} + B_{0} \partial_{j} B_{k}^{\dagger})
    \bar{q} \gamma_i q$  & + & + & 1 & $L=2$, $S=2$  \\ [5pt]
    V$9_-$ & $(1 / 2\Lambda^2) \imath \epsilon^{i 0 j k} (B_{0}^{\dagger} \partial_{j} B_{k} - B_{0} \partial_{j} B_{k}^{\dagger})
    \bar{q} \gamma_i q$  & - & + & 1 & $L=0,2$, $S=1$ \\
     & $(1 / 2\Lambda^2) \imath \epsilon^{i j 0 k} (B_{j}^{\dagger} \partial_{0} B_{k} - B_{j} \partial_{0} B_{k}^{\dagger})
    \bar{q} \gamma_i q$  & - & + & 1 & $L=0,2 $, $S=1$  \\ [5pt]
    V$10_+$ & $(1 / 2\Lambda^2) \epsilon^{i 0jk} (B_{0}^{\dagger} \partial_{j} B_{k} + B_{0} \partial_{j} B_{k}^{\dagger})
    \bar{q} \gamma_i \gamma^5 q$ & + & + & 1 & $L=2$, $S=2$  \\ [5pt]
    V$10_-$ & $(1 / 2\Lambda^2) \epsilon^{i 0 j k } (B_{0}^{\dagger} \partial_{j} B_{k} - B_{0} \partial_{j} B_{k}^{\dagger})
    \bar{q} \gamma_i \gamma^5 q$ & - & + & 1 & $L=0,2$, $S=1$   \\
    & $(1 / 2\Lambda^2) \epsilon^{i j 0 k} (B_{j}^{\dagger} \partial_{0} B_{k} - B_{j} \partial_{0} B_{k}^{\dagger})
    \bar{q} \gamma_i \gamma^5 q$ & - & + & 1 & $L=0,2$, $S=1$  \\
  \hline \hline
  \end{tabular}
  \caption{The charge conjugation ($C_B$), parity ($P_B$), and total angular momentum ($J$) quantum numbers of the DM system,
  as well as possible orbital ($L$) and spin ($S$) angular momenta of the DM state.  Only nonvanishing or non-negligible terms in each operator are shown.}
  \label{Table:CPJLS}
\end{table}

Using Table~\ref{Table:CPJLS}, one can first write the DM final state as a linear combination of states
in the $| L, S_{tot}, J, J_z \rangle$ basis (where $S_{tot}$ is the total spin of the DM
system),
then rewrite the state in the $| L, L_z, S_{tot}, S_{{tot}_z} \rangle$ basis, and finally rewrite the state in the $| L, L_z, S_1, S_{1z}, S_2, S_{2z} \rangle$ basis (where $S_1$ and
$S_2$ are the spins of each of the two dark matter particles.).  The matrix element then receives a factor
$E/m_B$ enhancement for each DM particle with spin projection $S_{(1,2)z} =0$.  The details of this
derivation are provided in Appendix A, and we summarize the energy dependence of the
leading term in Table~\ref{Table:Suppressions}.
\begin{table}[h!]
\begin{tabular}{l c c }
  \hline \hline
  Operators & Dimension enhancement & Polarization enhancement \\
  \hline
  V$1$, V2, V5, V6 & $E/\Lambda$ & $(E/m_B)^2$ \\
  V$3$, V4, V$7_-$, V$8_-$ & $(E/\Lambda)^2$ & $(E/m_B)^2$ \\
  V$7_+$, V$8_+$, V$9_\pm$, V$10_\pm$ & $(E/\Lambda)^2$ & $E/m_B$ \\
  \hline \hline
\end{tabular}
\caption{The energy enhancement factors in the matrix element for the process $\bar q q \rightarrow B^\dagger B$. }
\label{Table:Suppressions}
\end{table}
While it may be possible that the coefficient of the leading term
experiences an accidental cancellation, explicit calculation of the
matrix elements indicates that this is not the case.   The squared matrix
elements are listed in Appendix B.

\section{Constraints from unitarity}

If the scale of the new physics mediating the dark matter--quark interaction
is sufficiently light, the contact-operator approximation
will break down.  In simple models with a single mediator,
the energy suppression scale $\Lambda$ of the contact
operator is generally larger than the mass scale of the mediator,
implying that one should not trust the contact-operator approximation
for processes where the DM system center-of-mass energy
is larger than $\Lambda$.  But in more complicated models with large
numbers of mediators, it is not clear that the mediator mass must be
smaller than $\Lambda$.  More generally, the tightest constraint one
can rigorously impose is that the dark matter production matrix element
satisfy unitarity when evaluated at the energy of the hard process.

To impose this constraint, we follow the formalism and notation of~\cite{Endo:2014mja}.  
We thus consider the matrix element for the on-shell process $\bar q q \rightarrow B^\dagger B$.
If the initial state is a helicity eigenstate, then the fundamental unitarity constraint can be written as
\begin{eqnarray}
\sum_f \beta_i \beta_f |T^j_{i \rightarrow f}|^2  &\leq& 1\,,
\label{eqn:Unitarity}
\end{eqnarray}
where we have expanded the matrix element ${\cal M}_{i \rightarrow f}$ in Wigner $d$-functions as
\begin{eqnarray}
{\cal M}_{i \rightarrow f}(\theta) &=& 8\pi \sum_{j=0}^\infty (2j+1) T^j_{i \rightarrow f} d^j_{\lambda_f \lambda_i}\,.
\end{eqnarray}
Here, $\lambda_{i,f}$ are the total helicities of the initial and final states,
respectively, $j$ is the total angular momentum of the state, and $\theta$ is the scattering angle.
The Wigner $d$-functions $d^j_{\lambda_f \lambda_i}$ which are relevant here are given by
\begin{eqnarray}
d^0_{0,0} &=& 1 ,
\nonumber\\
d^1_{1,1} = d^1_{-1,-1} &=& {1 + \cos \theta \over 2}\,,
\nonumber\\
d^1_{1,0} = - d^1_{-1,0} =d^1_{0,-1} = - d^1_{0, 1} &=& -{\sin \theta \over \sqrt{2} }\,,
\nonumber\\
d^1_{1,-1} = d^1_{-1,1} &=& {1 - \cos \theta \over 2}\,,
\nonumber\\
d^1_{0,0} &=& \cos \theta\,.
\end{eqnarray}
They satisfy
\begin{subequations}
\begin{eqnarray}
\int_{-1}^1 d\cos \theta \, d^j_{\lambda' \lambda} d^{j'}_{\lambda' \lambda} &=& {2 \over 2j+1} \delta_{j j'} ,
\\
d^j_{\lambda \lambda'} (\theta =0) &=& \delta_{\lambda \lambda'}.
\end{eqnarray}
\end{subequations}

\begin{table}[t]
	{\renewcommand{\arraystretch}{3}
	\begin{tabular*}{0.75\textwidth}{ @{\extracolsep{\fill} } l  c c}
	\hline
	\hline
	Operator &  Constraint & Benchmark $\Lambda_{\text{min}}$ (TeV) \\ \hline
		V$1$,~V$2$ & \( \displaystyle \frac{E \sqrt{E^2 - m_B^2}}{16 \pi^2  \Lambda^2}\left( 3+\frac{4E^2}{m_B^4}(E^2 - m_B^2) \right) \leq 1 \) &  $1.59 \times 10^5$\\
		V$3$,~V$4$ & \( \displaystyle \frac{E (E^2 - m_B^2)^{3/2}}{72 \pi^2 \Lambda^4} \left( 3+\frac{4E^2}{m_B^4}(E^2 - m_B^2) \right) \leq 1 \) &  $274$\\
		V$5$,~V$6$ & \( \displaystyle \frac{E\sqrt{E^2-m_B^2}}{72 \pi^2  \Lambda^2} \left(\frac{4E^2}{m_B^2}+\frac{2E^2}{m_B^4}(E^2- m_B^2) -1  \right) \leq 1 \) &  $5.31 \times 10^4$\\
		V$7_+$,~V$8_+$ & \( \displaystyle \frac{E^3 (E^2 - m_B^2)^{3/2}}{18 \pi^2  m_B^2 \Lambda^4} \leq 1 \) & 8.66 \\
		V$9_+$,~V$10_+$ & \( \displaystyle \frac{E (E^2 - m_B^2)^{5/2}}{18 \pi^2 m_B^2 \Lambda^4} \leq 1 \) & 8.66 \\
		V$7_-$,~V$8_-$ & \( \displaystyle \frac{ E^3 (E^2 - m_B^2)^{3/2}}{18 \pi^2  m_B^2 \Lambda^4}  \left( 1 +  \frac{E^2}{m_B^2}  \right) \leq 1 \) & 274\\
		V$9_-$,~V$10_-$ & \( \displaystyle \frac{E^3 (E^2 - m_B^2)^{1/2}}{32 \pi^2   \Lambda^4}  \left( 1 + 2 \frac{E^2}{m_B^2}  \right) \leq 1 \) &  8.66\\[5pt]
	\hline
	\hline
	\end{tabular*}}
	\caption{Unitarity constraints on the energy $E$ of a dark matter particle in the center-of-mass frame of the $\bar{q}q \rightarrow B^{\dagger}B$ interaction.  These constraints can be rephrased in terms of a maximum $B^{\dagger} B$ invariant mass, which is then applied during event generation to get conservative collider sensitivities.  We have also included the minimum value of $\Lambda$  from these constraints for each operator, using the benchmark values of $E=1 \tev$ and $m_B=1 \gev$.}
	\label{Table:UnitarityBounds}
\end{table}

The unitarity constraints (on the $\bar{q}q \rightarrow B^{\dagger} B$ process) shown in Table~\ref{Table:UnitarityBounds} are obtained by explicitly evaluating Eq.~\eqref{eqn:Unitarity} for each of our 14 operators.  We also give the minimum value for $\Lambda$ allowed by these constraints, using the benchmark values of $E=1\tev$ and $m_B=1\gev$.  Our analysis will apply these constraints on an event-by-event basis: If we find that the LHC is sensitive to a particular value of $\Lambda$, the events used to establish this sensitivity had dark matter energies satisfying these constraints.  
Note that Eq.~\eqref{eqn:Unitarity} provides
constraints for each initial quark-antiquark state for which the matrix element is nontrivial.
For each of the operators there are at most two initial helicity eigenstates
that are relevant.  But for each operator with multiple possible initial helicity eigenstates,
the constraints arising from each of those possible initial states are degenerate.

In fact, because an energetic monojet is emitted, the actual hard process is 
$2 \rightarrow 3$ with an off-shell intermediate 
(anti)quark.  The unitarity bound 
above constrains the matrix element for the subprocess $\bar q q \rightarrow XX$, where the quark 
and antiquark are on shell.  However, for the energies and cuts relevant for the LHC analysis, 
the virtuality of the quark is a subleading effect, and a correct accounting for this virtuality results 
in only a small change in the unitarity constraint.  One could also use the unitarity condition to directly constrain 
the matrix element for the full $2 \rightarrow 3$ hard process, but this constraint is weaker than that arising 
from applying the unitarity condition to the $2 \rightarrow 2$ subprocess.  A more detailed discussion of these issues 
can be found in~\cite{Endo:2014mja}.

\section{LHC bounds}

We consider a search for monojet signatures arising from the process $pp \rightarrow B^\dagger B + \text{jets}$, assuming that
dark matter-quark interactions arise from any of the 14 contact operators shown in Table~\ref{Table:Operators}.
We assume a universal coupling to up and down quarks, and no coupling to the heavier generations.
Signal and SM background events are generated using the \textsc{MadGraph}/\textsc{Pythia}/\textsc{Delphes} simulation chain.
The 14 operators
are input into \textsc{MadGraph5}, and simulated $p p \rightarrow B^\dagger B j$ events for each operator are
generated using \textsc{MadEvent}~\cite{Alwall:2014hca}.  These events are then showered and hadronized using
\textsc{Pythia}-6.4~\cite{Sjostrand:2006za},
and event detection at ATLAS is simulated using \textsc{Delphes}-3 \cite{Ovyn:2009tx}.

We impose the unitarity constraints in Table~\ref{Table:UnitarityBounds} on an event-by-event basis~\cite{Racco:2015dxa}.
For each generated event for which the $\bar q q \rightarrow B^\dagger B$ matrix element satisfies
Eq.~\eqref{eqn:Unitarity}, the contact-operator approximation can provide an adequate description
of the physics.  For events where Eq.~\eqref{eqn:Unitarity} is not satisfied, some new physics
must come into play; to be conservative we simply
reject those events.  Note that the unitarity constraint for a specific choice of DM mass and $\Lambda$
is entirely a function of the invariant mass of the final state dark matter system; since the final states
of the selected events are different from the final states of the rejected events, interference effects are automatically
removed.

However, it is important to point out that for events that satisfy
Eq.~\eqref{eqn:Unitarity}, one only knows that the contact approximation is consistent with unitarity.
Although it is not necessary for there to be any additional non-negligible new physics, any particular model
may exhibit new physics not captured by the contact-operator approximation which is relevant at energies below the limit at which unitarity is violated.
In fact, it might be surprising if new physics fixed unitarity for processes
at energies where the contact approximation would slightly violate unitarity, but had only a negligible
impact even at slightly lower energies where the contact approximation would satisfy unitarity.

We also note that the unitarity analysis of~\cite{Endo:2014mja} and used in this work assumes that the incoming states
are in a helicity eigenstate.  However, for actual events at the LHC, the incoming partons of the hard process will
not generally be in a helicity eigenstate, but rather in a linear combination of helicity eigenstates.
To generalize the unitarity analysis to a generic initial state is beyond the scope of this work, but
we expect the maximum invariant mass for which unitarity is satisfied to change by at most
an ${\cal O}(\sqrt{2})$ factor.  Thus, although we have made largely conservative approximations, the
above caveats suggest that our limits may be uncertain by factors $\sim {\cal O}(\sqrt{2})$.

\subsection{Constraints from the $8\tev$ LHC}
The ATLAS Collaboration has performed a search \cite{Aad:2015zva} for the monojet signal using $20.3 \invfb$ of data at a center-of-mass energy of $\sqrt{s}=8 \tev$.  The main event selection criteria are the requirement of at least one jet with transverse momentum $p_T > 120 \gev$, several jet isolation and quality-control requirements, and a large amount of missing energy.  ATLAS chose several \textit{signal regions}, defined by differing amounts of minimum missing energy, in order to tune their analysis to various sources of new physics.  In this work, we will use their signal region SR4, requiring missing energy $\slashed{E}_T>300 \gev$.

The ATLAS analysis specifies the full set of triggers, jet candidate requirements, preselection cuts, and final signal-region missing-energy cuts performed on the $20.3 \invfb$ data set.  For this analysis, we use a slightly simplified subset of these cuts which are appropriate for our Monte-Carlo-generated event simulation chain and which capture the primary features of the ATLAS selection.  First, an initial cut of $\slashed{E}_T > 250 \gev$ is applied at the parton level in order to decrease the event veto rate and increase statistics after final cuts.  The detector level events are then required to satisfy the selection criteria given in Table~\ref{Table:Cuts8TeV}.  We have tested these cuts by applying them to $Z\rightarrow \nu \nu$, $W \rightarrow \nu l$, and $W \rightarrow \nu \tau $ background events produced using our same \textsc{MadGraph}/\textsc{Pythia}/\textsc{Delphes} simulation chain, and find that our predicted event rates match the ATLAS event rates to within $\sim 5\%$.

\begin{table}[t]
  \begin{tabular}{ l c }
	\hline
	\hline
	Jet reconstruction: & anti-$k_T$, using $R=0.4$ \\
	Jet definition: &  $p_{T}>30\gev$ and $|\eta|<4.5$	  \\
	Lepton veto: & electrons: $p_{T}>7\gev$ and $|\eta|<2.47$ \\
				    & muons: $p_{T}>7\gev$ and $|\eta|<2.5$ \\
	Leading jet: & $p_{Tj_1}>120\gev$ and $|\eta|<2$  \\
				 & $p_{Tj_1}/\slashed{E}_T > 0.5 $ \\
	Separation (all jets): & $\Delta \phi (p_{Tj},\slashed{E}_{T}) > 1.0$ \\
	Missing energy: & $\slashed{E}_T >300\gev$ \\
	\hline
	\hline
  \end{tabular}
  \caption{Monojet selection cuts for the $\sqrt{s}=8\tev$ LHC analysis.
  $\Delta \phi$ is the angular separation between the selected jet and the missing transverse
  momentum, $R$ is the radius parameter used in the anti-$k_T$ jet clustering algorithm~\cite{Cacciari:2008gp},
  and $\eta$ is the pseudorapidity.
  } \label{Table:Cuts8TeV}
\end{table}

Using the SR4 signal region, ATLAS is able to exclude at the 95\% C.L.\ any new-physics source of monojet events which gives rise to a cross section of $51 \fb$ or greater.  We use this constraint to bound the new physics scales $\Lambda$ for each of the 14 vector DM contact operators.
We apply the set of kinematical cuts in Table~\ref{Table:Cuts8TeV} to find the total event rate (or, the total cross section after cuts $\sigma_{\text{DM}}$) for a given $\Lambda$.

\begin{figure}
\includegraphics[width=0.75\textwidth]{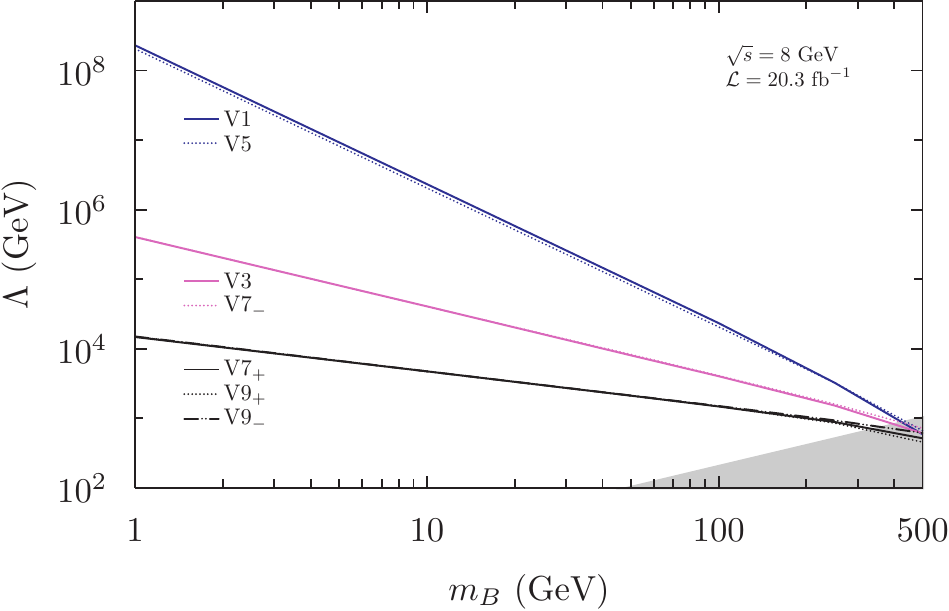}
\caption{\label{fig:Exclusions8TeV20invfb} The ATLAS 95\% C.L.\ exclusion bounds on the vector DM contact-operator scale $\Lambda$, using the $20.3 \invfb$ data set at $\sqrt{s}=8 \tev$.  All even-numbered operators (V2, V4, \ldots)\! are visually indistinguishable from their odd-numbered operator counterparts, and thus are not.
The shaded region at $\Lambda < 2 m_B$ represents the regime where the effective-operator description is naively expected to break down.  Note that all events used to establish these sensitivities were required to satisfy the unitarity constraints on four-point interaction energies.}
\end{figure}

To impose unitarity constraints, as discussed above, we apply at an event-by-event level  a cut on the maximum DM invariant mass, or equivalently, on the center-of-mass energy of the underlying DM-SM four-point interaction $\sqrt{s_{\text{DM}}}$.
Applying this cut safely underestimates the total event rate while excluding events from regions of phase space where unknown high-energy physics
is required by unitarity.
With this invariant mass cut in place, we tune $\Lambda$ so that the total event rate after all cuts corresponds to the new-physics cross section excluded by ATLAS, $\sigma_{\text{DM}} = 51 \fb$.  This provides us with 95\% C.L. exclusion bounds on $\Lambda$ for each operator, and over a range of DM masses $m_B$; these exclusion bounds are shown in Fig.~\ref{fig:Exclusions8TeV20invfb}.
We note here that the exclusion bounds for the even-numbered operators (V2, V4, \ldots)\! are visually indistinguishable from their odd-numbered counterparts over the mass range of Fig.~\ref{fig:Exclusions8TeV20invfb}, and hence are not included in the figure.

Operators V3, V5, V6, V7, and V9 would also permit the decay of a $J^{PC}= 1^{--}$ quarkonium state to $B^\dagger B$, if kinematically allowed.
Constraints on these operators from bounds on the invisible decay rate of $\Upsilon (1S)$ were found in~\cite{Fernandez:2014eja}, in the case
where dark matter couples to $b$-quarks.  We can compare those results (assuming a coupling to $b$-quarks) to the LHC sensitivity found above
(assuming instead a coupling to first-generation quarks).  In the region of overlap, the LHC reach in $\Lambda$ is at least a few orders of magnitude larger
than that obtained from current bounds on invisible quarkonium decay.

\subsection{Sensitivity of the $14 \tev$ LHC}
With the recent commencement of the high-energy LHC run, a new data set at increased center-of-mass energy will soon be in hand.  We now calculate the expected sensitivity of this new data set to spin-1 dark matter, coupling to Standard Model quarks through these 14 operators.  In order to best compare with the current exclusions found above, we calculate the expected 95\% C.L. exclusion bounds under the assumption that future data sets do not see any events above the expected background.  For this analysis, we will assume that the LHC Collaboration will soon increase beam energies to their full design energies for a total center-of-mass energy of $\sqrt{s}=14 \tev$, though these results should still be insightful if the LHC remains operating at $\sqrt{s}=13 \tev$.

To calculate exclusion bounds for the $14 \tev$ run, we must first know the event rate of SM irreducible background events at this energy.  We estimate this event rate by simulating the processes $Z\rightarrow \nu \nu$, $W \rightarrow \nu l$, and $W \rightarrow \nu \tau $ using the same Monte Carlo simulation chain and selection cuts used to calculate the signal event rate.  According to previous monojet analyses, including the ATLAS $20.3 \invfb$ analysis~\cite{Aad:2015zva}, these three processes contribute $\sim 95\%$ of the total irreducible background to the monojet signal, so we neglect other sources such as $t \bar{t}$ and single top production.

The simulation chain is identical to that used for the $8 \tev$ data set described above with adjustments to the parton-level and detector-level jet definitions and cuts.  These adjustments are motivated by an ATLAS Collaboration study of monojet searches at $14 \tev$~\cite{ATLAS14TeVStudy}.  Jets are defined to have a higher minimum transverse momentum of $p_T > 50 \gev$ in order to mitigate the higher pileup expected at this increased energy.  Additionally, all jets are required to be more central, with $|\eta| < 3.6$, as jet reconstruction algorithm performance is not well understood in the forward regions.  The leading-jet minimum $p_T$ and minimum missing energy are both increased to reflect the increase in available energy to $p_{Tj1} > 300 \gev$ and $\slashed{E}_T > 600 \gev$, and the parton-level cut on missing energy is correspondingly increased to $\slashed{E}_T > 500 \gev$ to increase statistics.  All final selection cuts are shown in Table~\ref{Table:Cuts14TeV}.

\begin{table}[t]
  \begin{tabular}{ l c }
	\hline
	\hline
	Jet reconstruction: & anti-$k_T$, using $R=0.4$ \\
	Jet definition: &  $p_{T}>50\gev$ and $|\eta|<3.6$	  \\
	Lepton veto: & electrons: $p_{T}>7\gev$ and $|\eta|<2.47$ \\
				    & muons: $p_{T}>7\gev$ and $|\eta|<2.5$ \\
	Leading jet: & $p_{Tj_1}>300\gev$ and $|\eta|<2.0$  \\
				 & $p_{Tj_1}/\slashed{E}_T > 0.5 $ \\
	Separation (all jets): & $\Delta \phi (p_{Tj},\slashed{E}_{T}) > 1.0$ \\
	Missing energy: & $\slashed{E}_T >600\gev$ \\
	\hline
	\hline
  \end{tabular}
  \caption{Monojet event selection cuts for the $\sqrt{s}=14 \tev$ LHC analysis.  These cuts are similar to those used for the $8 \tev$ LHC, with adjustments to jet definition, leading-jet momentum, and missing energy, motivated by the higher energy and pileup at 14 TeV \cite{ATLAS14TeVStudy}.}
  \label{Table:Cuts14TeV}
\end{table}

On applying these cuts to the generated background events, we find the cross sections for the three dominant background processes, as shown in Table~\ref{Table:Backgrounds14TeV}.  These event rates are represented by $\tilde{\sigma}_{\text{SM}} \times \epsilon $, where $\tilde{\sigma}_{\text{SM}}$ represents the cross section for the process as calculated from the \textsc{MadGraph}/\textsc{Pythia}/\textsc{Delphes} simulation chain for our choice of parton-level cuts, and $\epsilon$ represents the efficiency of the selection cuts given in Table~\ref{Table:Cuts14TeV}.  The product $\tilde{\sigma}_{\text{SM}} \times \epsilon $
is the physically meaningful quantity, as both $\tilde{\sigma}_{\text{SM}}$ and $\epsilon$ depend on our choice of parton-level cuts.

\begin{table}[t]
	\begin{tabular*}{0.30\textwidth}{ @{\extracolsep{\fill} } l c}
	\hline
	\hline
		& $~~~~~\tilde{\sigma}_{\text{SM}} \times \epsilon ~(\text{fb})$ \\ \hline
		$Z \rightarrow \nu \nu$ & $79.93$\\
		$W \rightarrow \tau \nu_\tau$ & $16.99$ \\
		$W \rightarrow l \nu_l$ & $10.51$ \\
		\textbf{Total} & $\mathbf{107.4}$ \\
	\hline
	\hline
	\end{tabular*}
	\caption{Background cross sections at the $\sqrt{s}=14 \tev$ LHC for the three dominant background processes using the cuts given in Table~\ref{Table:Cuts14TeV}.  As determined by previous monojet analyses, these three processes contribute $\sim 95\%$ of the total SM irreducible background, so we neglect other sources in our background estimation.}
	\label{Table:Backgrounds14TeV}
\end{table}

Using these results, we calculate the total new-physics cross section that can be excluded by the $14 \tev$ LHC for a given integrated luminosity.  The $95\%$ C.L.\ expected exclusion bound is given by
\begin{equation}
\chi^2 \equiv \frac{N_{\text{DM}}(\Lambda)^2}{N_{\text{SM}}+N_{\text{DM}}(\Lambda)+\Sigma_{\text{SM}}^2} = 3.84\,,
\end{equation}
where $N_{\text{SM}}$ and $N_{\text{DM}}$ are the number of expected background and signal events, respectively, and $\Sigma_{\text{SM}}$ is the systematic uncertainty in the number of background events.  We can rewrite the above condition in terms of integrated luminosity as
\begin{equation}
 \frac{\sigma_{\text{DM}}(\Lambda)^2 \cdot {\cal L}}{\sigma_{\text{SM}}+\sigma_{\text{DM}}(\Lambda)+\delta_{\text{SM}}^2 \cdot \sigma_{\text{SM}}^2 \cdot {\cal L}} = 3.84\,, \label{eq:95CLExclusion}
\end{equation}
where $\sigma \equiv \tilde{\sigma} \times \epsilon$ is the cross section after cuts, and $\delta_{\text{SM}}$ is the percent systematic uncertainty in the number of background events.   As determined by the ATLAS Collaboration, the uncertainty in background events at $8 \tev$ ranges from $\sim 2\%$ to  $10\%$ (see, for instance, Tables~4 and 5 of \cite{Aad:2015zva}), so for this analysis we take $\delta_{\text{SM}}=0.05$.  We note here that although the LHC sensitivity to each individual operator will depend on this uncertainty, our plots which overlay all vector operators extend over such a large range that small variations of $\delta_{\text{SM}}$ will appear negligible.  For this analysis we take
 ${\cal L} = 100 \invfb$, which is the approximate integrated luminosity expected to be collected by the end of the first LHC run at $14 \tev$.  For these values of ${\cal L}$, $\delta_{\text{SM}}$, and $\sigma_{\text{SM}}$ (as determined above and displayed in Table \ref{Table:Backgrounds14TeV}), we find that the LHC at $14 \tev$ is able to exclude at the 95\% C.L.\ a new-physics event rate of $\sigma_{\text{DM}}(\Lambda)=10.74 \fb$.  This sensitivity is dominated by systematic uncertainties, and thus roughly scales linearly with $\delta_{\text{SM}}$.

\begin{figure}
\includegraphics[width=0.75\textwidth]{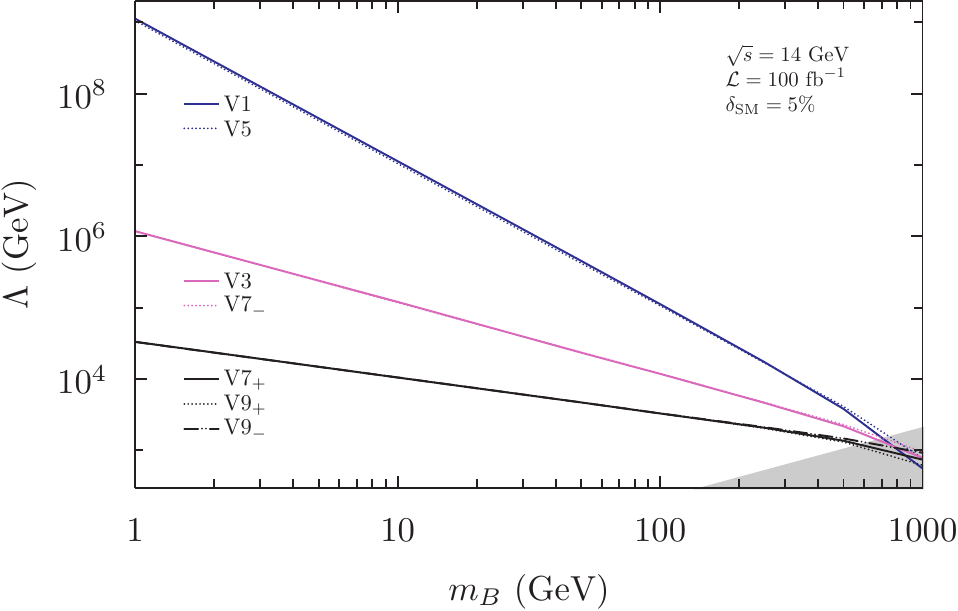}
\caption{\label{fig:Exclusions14TeV100invfb} The expected ATLAS 95\% C.L.\ exclusion bounds on the vector DM contact-operator scale $\Lambda$, using $100 \invfb$ of data at $\sqrt{s}=14 \tev$.  All even-numbered operators (V2, V4, \ldots) are visually indistinguishable from the odd-numbered operators shown.
The shaded region at $\Lambda < 2 m_B$ represents the regime where the effective operator description is naively expected to break down.}
\end{figure}

Analysis of the signal events for each of the 14 contact operators proceeds in the same way as before, where $\Lambda$ is tuned such that the total event rate after cuts, including the cut on DM invariant mass constrained by unitarity, is such that $\sigma_{\text{DM}}(\Lambda)=10.74 \fb$.  The exclusion bounds on $\Lambda$ over a range of DM masses are shown in Fig.~\ref{fig:Exclusions14TeV100invfb}.  Once again, only the odd-numbered operators are shown, as the even-numbered operators are visually indistinguishable from their odd-numbered counterparts on this plot.

\subsection{Discussion of sensitivities}
\label{sec:Discussion}

The results shown in Figs.~\ref{fig:Exclusions8TeV20invfb} and~\ref{fig:Exclusions14TeV100invfb} encode the enhancements each operator receives from both operator dimension and longitudinal polarizations, and the features in these plots can be fully understood by a simple accounting of these enhancements.  From Table~\ref{Table:Suppressions}, we see that there are three subsets of operators that are defined by how they scale: operators V$(1,2,5,6)$ scale as $E^3 /(\Lambda m_{B}^{2})$, operators V$(3,4,7_-,8_-)$ scale as $E^4 /( \Lambda^2 m_B^2)$, and operators V$(7_+,8_+,9_{\pm},10_{\pm})$ scale as $E^3 /( \Lambda^2 m_B)$.  Thus depending on the operator there will be three different ways in which the $pp \rightarrow B^{\dagger} B$ cross section, and thus the collider sensitivity to $\Lambda$, scales.  This is shown in Table~\ref{Table:SensitivityScaling}, and the behavior can clearly be seen in the LHC exclusion bounds shown in Figs.~\ref{fig:Exclusions8TeV20invfb} and \ref{fig:Exclusions14TeV100invfb}.

\begin{table}
	\begin{tabular*}{0.5\textwidth}{ @{\extracolsep{\fill} } l c c}
	\hline
	\hline
		& $\sigma \propto $ & $\Lambda \propto $ \\ \hline
		V$1$, V$2$, V$5$, V$6$ & $\displaystyle \frac{E^6}{\Lambda^2 m_{B}^{4}}$ & $\displaystyle E \left( \sqrt{\frac{E}{m_B}} \right)^4$  \\[10pt]
		V$3$, V$4$, V$7_{-}$, V$8_{-}$ & $\displaystyle \frac{E^8}{\Lambda^4 m_{B}^{4}}$ & $\displaystyle E \left( \sqrt{\frac{E}{m_B}} \right)^2$ \\[10pt]
		V$7_{+}$, V$8_{+}$, V$9_{\pm}$, V$10_{\pm}$ & $\displaystyle \frac{E^6}{\Lambda^4 m_{B}^{2}}$ & $\displaystyle E \sqrt{\frac{E}{m_B}}$ \\[10pt]
	\hline
	\hline
	\end{tabular*}
	\caption{Scaling of the $\bar{q}q \rightarrow B^{\dagger} B$ cross section and overall collider sensitivity to $\Lambda$ based on the enhancements from operator dimension and longitudinal polarizations given in Table~\ref{Table:Suppressions}.  We see that the three subsets of operators should scale differently, which is indeed the behavior seen in Figs.~\ref{fig:Exclusions8TeV20invfb} and \ref{fig:Exclusions14TeV100invfb}.}
	\label{Table:SensitivityScaling}
\end{table}

Furthermore, the overall \textit{magnitude} of the exclusion limits in Figs.~\ref{fig:Exclusions8TeV20invfb} and \ref{fig:Exclusions14TeV100invfb} matches what we expect.  For instance, at the LHC we expect the center-of-mass energy of these quark-DM interactions to be ${\cal O}(1 \tev)$; given a DM mass of $m_B = 1\gev$, we then expect $\sqrt{E / m_B} \sim 30$.  According to Table~\ref{Table:SensitivityScaling} the sensitivities to $\Lambda$ for the three different subsets of operators should then, for $m_B=1$~GeV, stand approximately in the ratio $1: 30 : 30000$.  This is indeed what is seen in the figures.  Thus it becomes clear that collider sensitivity to vector-DM production can receive an enhancement of \textit{several orders of magnitude} at the LHC simply due to the presence of the longitudinal polarization mode.  We now study how these collider bounds on $\Lambda$ translate to constraints on scattering cross sections at direct-detection experiments.

\subsection{LHC bounds on velocity-independent scattering}

Operators V$1$ and V$3$ permit velocity-independent spin-independent (SI) scattering, while operators
V$5$ and V$10_+$ permit velocity-independent spin-dependent (SD) scattering (see, for example,~\cite{Kumar:2013iva}).
For these operators, LHC constraints on the energy scale of the operator ($\Lambda$) can be expressed as
constraints on the DM--nucleon scattering cross section.

The DM--nucleon scattering cross sections may be written as
\begin{eqnarray}
\sigma^N &=& {\mu_N^2 \over 16\pi m_B^2 m_N^2} \left({1 \over 6} \sum_{spins} |{\cal M}|^2 \right)\, ,
\end{eqnarray}
where $\mu_N$ is the reduced mass of the DM--nucleon system, $m_N$ is the nucleon mass, and
${\cal M}$ is the scattering matrix element.  Assuming isospin-invariant couplings to first-generation
quarks, the scattering cross sections can be written as (see also~\cite{Fernandez:2014eja})
\begin{subequations}
\begin{eqnarray}
\sigmaSI^{N(\text{V1})} &=& {\mu_N^2 \over 4\pi m_B^2 \Lambda^2 } \left(B_u^{N(s)} + B_d^{N(s)}\right)^2\, ,
\\
\sigmaSI^{N(\text{V3})} &=& {\mu_N^2 \over 4\pi \Lambda^4 } \left(B_u^{N(v)} + B_d^{N(v)}\right)^2 \,,
\\
\sigmaSD^{N(\text{V5})} &=& {\mu_N^2 \over 2\pi m_B^2 \Lambda^2 } \left(B_u^{N(t)} + B_d^{N(t)}\right)^2 \,,
\\
\sigmaSD^{N(\text{V10}_+)} &=& {\mu_N^2 \over 2\pi \Lambda^4 } \left(B_u^{N(pv)} + B_d^{N(pv)}\right)^2 \,,
\end{eqnarray}
\end{subequations}
where $B_{u,d}^{N(s,v,pv,t)}$ are the nucleon form factors for $u$- and $d$-quarks for scalar,
vector, pseudovector, and tensor structures.  The vector nucleon form factors are fixed by
gauge invariance, and are given by
\begin{eqnarray}
B_u^{p(v)} = B_d^{n(v)} =2\,,  \qquad   B_u^{n(v)} = B_d^{p(v)} = 1\,.
\end{eqnarray}
The remaining nucleon form factors are subject to uncertainties related to the structure of the
nucleon. We  use the following
values as benchmarks~\cite{Ellis:2009ai,Fan:2010gt,Kelso:2014qja}:

\begin{align}
B_u^{p(s)} \sim B_d^{n(s)} &\sim 10\,,&    B_u^{n(s)} \sim B_d^{p(s)} &\sim  7\,,&
\nonumber\\
B_u^{p(pv)} \sim B_d^{n(pv)} &\sim 0.84\,,&     B_u^{n(pv)} \sim B_d^{p(pv)} &\sim -0.43\,,&
\nonumber\\
B_u^{p(t)} \sim B_d^{n(t)} &\sim 0.54\,,&     B_u^{n(t)} \sim B_d^{p(t)} &\sim  -0.23\,.&
\end{align}

\begin{figure}
\includegraphics[width=0.75\textwidth]{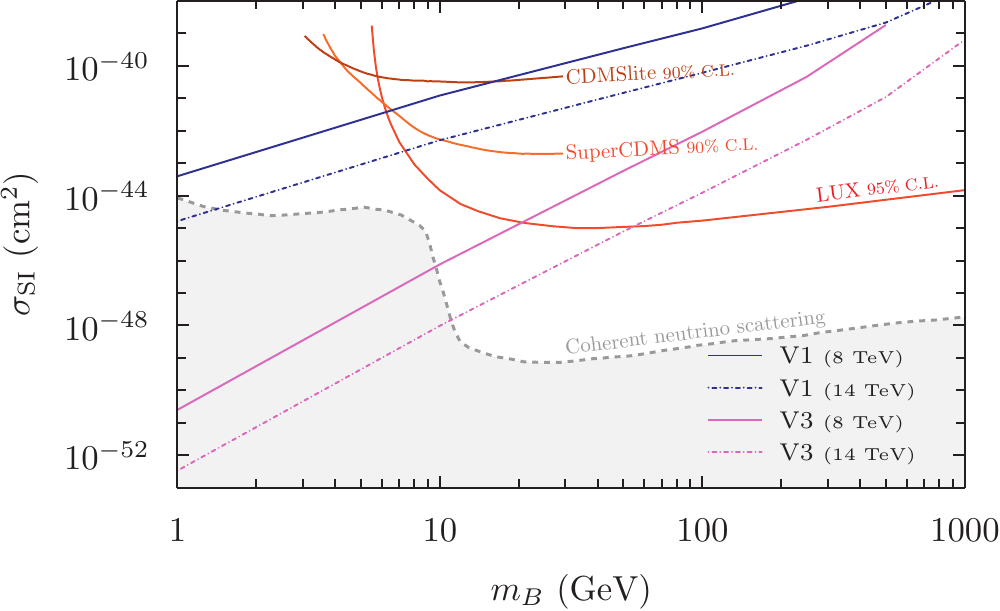}
\caption{\label{fig:CrossSectionBounds_SI} The ATLAS 95\% C.L.\ exclusion bounds on $\sigmaSI^{(N)\text{V1},\text{V3}}$,
using the $20.3 \invfb$ data set at $\sqrt{s}=8 \tev$, as well as the expected 95\% C.L.\ sensitivity curve for a $100 \invfb$ data set at $\sqrt{s}=14 \tev$.
Also plotted are exclusion bounds from SuperCDMS \cite{SuperCDMS}, CDMSlite \cite{CDMSlite}, and LUX \cite{Akerib:2013tjd}, as well as a representative minimum cross section below which neutrino--nucleus coherent scattering is significant~\cite{Billard:2013qya}.  }
\end{figure}

\begin{figure}[t]
\includegraphics[width=0.75\textwidth]{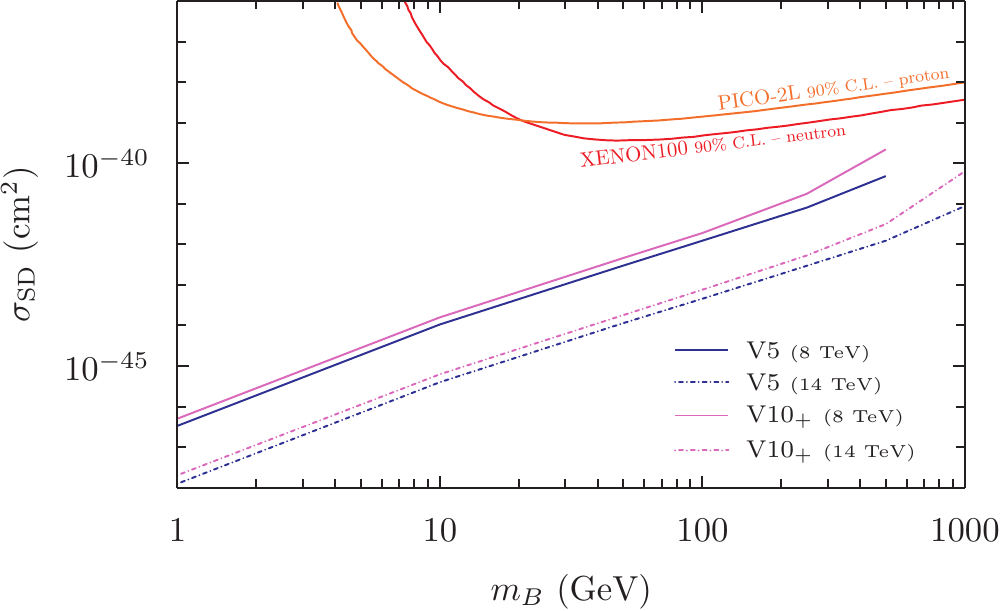}
\caption{\label{fig:CrossSectionBounds_SD} The ATLAS 95\% C.L.\ exclusion bounds on $\sigmaSD^{(N)\text{V5},\text{V10}}$
using the $20.3 \invfb$ data set at $\sqrt{s}=8 \tev$.  Also plotted are 95\% C.L.\ exclusion bounds from PICO-2L~\cite{Amole:2015lsj} and XENON100~\cite{Aprile:2012nq}.}
\end{figure}

We plot current LHC 95\%~C.L.\ bounds on $\sigmaSI$ and $\sigmaSD$ in Figs.~\ref{fig:CrossSectionBounds_SI}~and~\ref{fig:CrossSectionBounds_SD}, respectively.  We also plot current direct-detection limits from LUX \cite{Akerib:2013tjd}, SuperCDMS~\cite{SuperCDMS}, and CDMSlite~\cite{CDMSlite} for spin-independent scattering, and PICO-2L~\cite{Amole:2015lsj} and XENON100~\cite{Aprile:2012nq} for spin-dependent scattering.  In addition, we plot the minimum spin-independent scattering cross section that can be probed while still having
an insignificant contribution of events from neutrino--nucleus coherent scattering~\cite{Billard:2013qya}.

For operators that produce spin-independent scattering, the LHC is only
competitive with direct-detection experiments at low mass.  Note that for operator V$1$, although the LHC
energy reach $\Lambda$ increases dramatically for small $m_B$, its sensitivity to $\sigmaSI$ does not increase as
dramatically because $\sigmaSI^{N(\text{V1})} \propto 1/m_B^{2}$.  Additionally, despite the fact that collider sensitivity to $\Lambda$
for V$1$ far exceeds the sensitivity for V$3$, collider bounds more tightly constrain $\sigmaSI$ for V$3$.

For spin-dependent scattering, however, the LHC sensitivity far exceeds that of current direct-detection experiments.  Additionally, the LHC is more sensitive to vector operators V$5$ and V$10_+$ than it is to fermionic operators which lead to SD scattering, such as $\bar{\chi} \gamma^{\mu} \gamma^5 \chi \ \bar{q} \gamma_{\mu} \gamma^5 q$ and $\bar{\chi}  \sigma^{\mu \nu} \chi \ \bar{q} \sigma_{\mu \nu}  q$.  This is also the case for vector and fermionic spin-independent operators.

\section{Conclusion}

We considered the sensitivity of the LHC to spin-1 dark matter that couples to first-generation quarks via a contact operator.  We found that a monojet search strategy can
probe contact operators with an energy suppression scale $\Lambda$ which can range from
${\cal O}(1-10^5)\tev$.
This large energy reach results from dark matter final states which include one or more longitudinally
polarized spin-1 particles, yielding large enhancements to the production matrix element.
In this analysis, we imposed conservative cuts to ensure that
we only consider regions of phase space for which the contact-operator approximation is consistent
with unitarity.  For operators that permit velocity-independent SD scattering, these bounds far exceed those obtainable from current direct-detection experiments, while for operators that permit
velocity-independent SI scattering, these bounds are comparable to those obtainable from
direct-detection experiments for light dark matter.

It is worth noting the implications of these results for future high-energy hadron colliders.  As indicated in
Table~\ref{Table:SensitivityScaling}, the sensitivity of monojet searches to spin-1 dark matter scales as a
high positive power of the characteristic energy of the collider; for operators V$(1,2,5,6)$, the sensitivity scales
as $\Lambda \propto E^3$ and an increase in collider energy is much more effective than an increase in luminosity.
Indeed, for these operators, the sensitivity of a $14 \tev$ run of the LHC is approximately an order of magnitude
greater than that of an $8 \tev$ run.  One might therefore expect that a future ${\cal O}(100) \tev$ hadron collider could
provide a sensitivity orders of magnitude greater than current bounds.


\begin{acknowledgments}
We are grateful to Patrick Stengel for useful discussions.
JK would like to thank CETUP* (Center for Theoretical Underground Physics and Related Areas),
for its hospitality and partial support during the 2015 Summer Program. DM thanks the Mainz Institute for Theoretical Physics and the Aspen Center for Physics (which is supported by National Science Foundation grant PHY-1066293) for their hospitality and partial support during the completion of this work.
JK is supported in part by NSF CAREER grant PHY-1250573.  DM is supported in part by
DOE grant DE-SC0010504.  DY is supported in part by DOE grant DE-FG02-13ER-41976.
\end{acknowledgments}

\appendix

\section{Energy enhancement due to longitudinal polarization}

The two-particle spin state in the $|S_1, S_2, S_{tot}, S_{{tot}_z} \rangle$ basis can be expressed
in the $|S_1 , S_{1z}; S_2, S_{2z} \rangle$ basis as follows: \begin{eqnarray}
|1, 1, 2, 2 \rangle &=& |1, 1; 1, 1 \rangle\, ,
\nonumber\\
|1, 1, 2, 1 \rangle &=& {1\over \sqrt{2}}|1, 0; 1, 1 \rangle + {1\over \sqrt{2}}|1, 1; 1, 0 \rangle\, ,
\nonumber\\
|1, 1, 2, 0 \rangle &=& {1\over \sqrt{6}}|1, -1; 1, 1 \rangle +{2\over \sqrt{6}}|1, 0; 1, 0 \rangle +{1\over \sqrt{6}}|1, 1; 1, -1 \rangle\, ,
\nonumber\\
|1, 1, 2, -1 \rangle &=& {1\over \sqrt{2}}|1, 0; 1, -1 \rangle + {1\over \sqrt{2}}|1, -1; 1, 0 \rangle\, ,
\nonumber\\
|1, 1, 2, -2 \rangle &=& |1, -1; 1, -1 \rangle \,,
\nonumber\\
|1, 1, 1, 1 \rangle &=& {1\over \sqrt{2}}|1, 0; 1, 1 \rangle - {1\over \sqrt{2}}|1, 1; 1, 0 \rangle \,,
\nonumber\\
|1, 1, 1, 0 \rangle &=& {1\over \sqrt{2}}|1, -1; 1, 1 \rangle - {1\over \sqrt{2}}|1, 1; 1, -1 \rangle\,,
\nonumber\\
|1, 1, 1, -1 \rangle &=& {1\over \sqrt{2}}|1, 0; 1, -1 \rangle - {1\over \sqrt{2}}|1, -1; 1, 0 \rangle \,,
\nonumber\\
|1, 1, 0, 0 \rangle &=& {1\over \sqrt{3}}|1, -1; 1, 1 \rangle -{1\over \sqrt{3}}|1, 0; 1, 0 \rangle +{1\over \sqrt{3}}|1, 1; 1, -1 \rangle\,.
\end{eqnarray}
Since the matrix element receives an $E/m_B$ enhancement factor for each longitudinal polarization,
we find the following relations between the DM total spin state
and the leading energy enhancement:
\begin{eqnarray}
(E/m_B)^2 &:& S_{tot}=2, S_{{tot}_z} =0; S_{tot}=0, S_{{tot}_z}=0 \,,
\nonumber\\
(E/m_B)^1 &:& S_{tot}=2, S_{{tot}_z} =\pm 1; S_{tot}=1, S_{{tot}_z} =\pm 1 \,,
\nonumber\\
(E/m_B)^0 &:& S_{tot}=2, S_{{tot}_z} =\pm 2; S_{tot}=1, S_{{tot}_z} = 0 \,.
\end{eqnarray}

The last step is simply to consider each $| L, S_{tot}, J, J_z \rangle $ final state
allowed by every term of every operator~\cite{Kumar:2013iva}, and expand in the basis
$| L, L_z, S_{tot}, S_{tot_{z}} \rangle $ via Clebsch-Gordan coefficients.  We then find the
leading energy enhancement allowed, yielding
\begin{eqnarray}
L=0, S_{tot}=0, J=0 &\rightarrow & (E/m_B)^2 \,,
\nonumber\\
L=2, S_{tot}=2, J=0 &\rightarrow & (E/m_B)^2 \,,
\nonumber\\
L=1, S_{tot}=0, J=1 &\rightarrow & (E/m_B)^2 \,,
\nonumber\\
L=0, S_{tot}=1, J=1 &\rightarrow & E/m_B \,,
\nonumber\\
L=1, S_{tot}=1, J=1 &\rightarrow & E/m_B\,,
\nonumber\\
L=2, S_{tot}=1, J=1 &\rightarrow & E/m_B \,,
\nonumber\\
L=2, S_{tot}=2, J=1 &\rightarrow & E/m_B \,.
\end{eqnarray}
Note that these expressions are sufficient to compute the leading energy enhancement for
each of the operators we consider.

\section{Squared matrix elements}

Here we list the  squared matrix elements (summed over spins and
polarizations) for the process $\bar q q \rightarrow B^\dagger B$.
We take $E$ to be the energy of a DM particle in the
center-of-mass frame of the dark matter system, so that $E^2 = k_q^2 + m_q^2 = k_B^2 + m_B^2$.
The angle between the axis of the quark-antiquark system and the axis of the dark matter
system is $\theta$.
In the limit $m_q \rightarrow 0$, we find
{\allowdisplaybreaks
\begin{align}
\sum_{spins}|{\cal M}_{\text{V1}}|^2 = \sum_{spins}|{\cal M}_{\text{V2}}|^2
&= \frac{8E^2}{\Lambda^2} \left( 3 + 4 \frac{\vec{k}^{2}_{B}}{m_{B}^{2}} \left (1+ \frac{\vec{k}^{2}_{B}}{m_{B}^{2}}\right )  \right) ,
\\
\sum_{spins}|{\cal M}_{\text{V3}}|^2 = \sum_{spins}|{\cal M}_{\text{V4}}|^2 &= \frac{8E^2 \vec{k}^{2}_{B} }{\Lambda^4} \sin ^2 \theta \left ( 3+ 4 \frac{\vec{k}^{2}_{B}}{m_{B}^{2}}  \left(1 +  \frac{\vec{k}^{2}_{B}}{m_{B}^{2}}  \right)  \right ) ,
\\
\sum_{spins}|{\cal M}_{\text{V5}}|^2 = \sum_{spins}|{\cal M}_{\text{V6}}|^2 & = \frac{16 E^2}{\Lambda^2}\left (1 + 2 \frac{\vec{k}^{2}_{B}}{m^{2}_{B}}\left( 1+ \frac{\vec{k}^{2}_{B}}{m^{2}_{B}} \cos^2 \theta \right ) \right) ,
\\
\sum_{spins}|{\cal M}_{\text{V7}_{+}}|^2 = \sum_{spins}|{\cal M}_{\text{V8}_{+}}|^2 &= \frac{16 E^4 }{\Lambda^4 } \frac{\vec{k}^{2}_{B}}{m^{2}_{B}} \left ( 1+\cos^2 \theta \right ) ,
\\
\sum_{spins}|{\cal M}_{\text{V9}_{+}}|^2 = \sum_{spins}|{\cal M}_{\text{V10}_{+}}|^2  &=  \frac{16 E^2 }{\Lambda^4 } \frac{ \vec{k}^{4}_{B}}{m^{2}_{B}} \left ( 1+\cos^2 \theta \right ).
\end{align}
} 
For the V$(7-10)_{-}$ operators, we find
\begin{align}
\sum_{spins}|{\cal M}_{\text{V7}_{-}}|^2 = \sum_{spins}|{\cal M}_{\text{V8}_{-}}|^2 &= \frac{16 E^4}{\Lambda^4} \frac{\vec{k}^{2}_{B}}{m^{2}_{B}}
\left( 1 + \cos^2 \theta + 2\frac{ E^2 }{ m^{2}_{B} }  \sin^2 \theta \right ) ,
\\
\sum_{spins}|{\cal M}_{\text{V9}_{-}}|^2 = \sum_{spins}|{\cal M}_{\text{V10}_{-}}|^2&= \frac{16 E^4}{\Lambda^4}   \left( \frac{E^2}{m^{2}_{B}}
\left( 1+ \cos^2 \theta \right) + \sin^2 \theta \right).
\end{align}

\section{Properties of operators V$(7-10)_-$}

Operators V$(7-10)_-$ were not discussed in~\cite{Kumar:2013iva}.  We repeat much
of that analysis for these operators, for completeness.
In Table~\ref{Table:PrimedAnnihMatrixElements}, we write the annihilation
matrix element factors that arise from the DM bilinears relevant for
operators V$(7-10)_-$.  We assume that the DM system is in the center-of-mass frame, with $B$ ($B^\dagger$) having spatial momentum $\overrightarrow{k}$ ($-\overrightarrow{k}$).
In Table~\ref{Table:PrimedScatteringMatrixElements}, we write the scattering
matrix element factors in the center-of-mass frame that arise from the same bilinears, where the incoming
DM particle has polarization vector $\epsilon$ and the outgoing particle
has polarization vector $\epsilon'$.
In Table~\ref{table:PrimedOpsMaster}, we indicate if the operator permits
DM $s$-wave annihilation, and indicate the factors of momentum transfer ($q$)
and of DM velocity perpendicular to momentum transfer ($v^\bot$) that suppress
the spin-independent and spin-dependent scattering cross sections (including if the factor arises
from the DM or SM bilinear).

\begin{table}[h]
\begin{center}
\begin{tabular}{l c}
  \hline \hline
  Bilinear & Annihilation matrix element \\
\hline
   $\epsilon^{0ijk}(B^\dagger_i \partial_j B_k -B_i \partial_j B^\dagger_k )$ &
    0 \\
  $\epsilon^{0ijk}(B^\dagger_j \partial_0 B_k - B_j \partial_0 B^\dagger_k)$ &
  $\imath E \epsilon^{ijk} (\epsilon_1^j \epsilon_2^k - \epsilon_2^j \epsilon_1^k)$ \\
  $-\epsilon_{0ijk}(B^\dagger_0 \partial_j B_k - B_0 \partial_j B^\dagger_k )$ &
  $\imath \epsilon_{ijk} k^j (\epsilon_2^0 \epsilon_1^k + \epsilon_2^k \epsilon_1^0) $ \\
 $-\epsilon_{kij0}(B^\dagger_k \partial_j B_0 - B_k \partial_j B^\dagger_0 )$ &
  $-\imath \epsilon_{ijk} k^j (\epsilon_2^k \epsilon_1^0 + \epsilon_2^0 \epsilon_1^k) $ \\
  $B^{\dagger \nu} \partial_\nu B_0 -B^\nu \partial_\nu B^\dagger_0 $ &
   0 \\
  $B^{\dagger \nu} \partial_\nu B_i - B^\nu \partial_\nu B^\dagger_i$  &
  $2\imath E (\epsilon_2^0 \epsilon_1^i - \epsilon_1^0 \epsilon_2^i) $ \\
  \hline \hline
\end{tabular}
\caption{The annihilation matrix element factors for spin-1 dark matter bilinears relevant for operators V$(7-10)_-$.}
\label{Table:PrimedAnnihMatrixElements}
\end{center}
\end{table}

\begin{table}[h]
\begin{center}
\begin{tabular}{l c}
  \hline \hline
  Bilinear & Scattering matrix element (SD)\\
\hline
$(B^\dagger_\nu \partial^\nu B_\mu - B_\nu \partial^\nu B^\dagger_\mu)$ &
$-\imath q^i (\epsilon'_i \epsilon_\mu + \epsilon_i \epsilon'_\mu)$  \\
$\epsilon^{0 \nu \rho \sigma} (B^\dagger_\nu \partial_\rho B_\sigma - B_\nu \partial_\rho B^\dagger_\sigma)$ &
$-2 \imath \epsilon^{ijk} q_i \epsilon_j \epsilon'_k $ \\
$\epsilon^{i \nu \rho \sigma} (B^\dagger_\nu \partial_\rho B_\sigma - B_\nu \partial_\rho B^\dagger_\sigma)$ &
0 \\
  \hline \hline
\end{tabular}
\caption{The scattering matrix element factors for spin-1 dark matter bilinears relevant for operators V$(7-10)_-$.}
\label{Table:PrimedScatteringMatrixElements}
\end{center}
\end{table}

\begin{table}[h]
\begin{center}
\scriptsize
\begin{tabular}{l c c c c }
\hline \hline
  Operator &  Structure & $\sigmaSI$ suppression & $\sigmaSD$ suppression &$s$-wave?\\
\hline
  V$7_-$ & $(B^\dagger_\nu \partial^\nu B_\mu - B_\nu \partial^\nu B^\dagger_\mu) \bar q \gamma^\mu q$  & $v^{\bot 2}$ (SM); $q^2$ (DM)
  & $q^2$ (SM); $q^2$ (DM)  & No \\
& & $q^2 v^{\bot 2}$ (DM) & & \\ [5pt]
  V$8_-$ & $(B^\dagger_\nu \partial^\nu B_\mu - B_\nu \partial^\nu B^\dagger_\mu) \bar q \gamma^\mu \gamma^5 q$ &
  $q^2 v^{\bot 2}$ (SM); $q^2$ (DM) & $q^2$ (DM) & No \\[5pt]
  V$9_-$ & $\epsilon^{\mu \nu \rho \sigma} (B^\dagger_\nu \partial_\rho B_\sigma - B_\nu \partial_\rho B^\dagger_\sigma) \bar q \gamma_\mu q$ &
  $q^2$ (DM) & $q^2 v^{\bot 2}$ (SM); $q^2$ (DM) & Yes \\[5pt]
  V$10_-$ & $\epsilon^{\mu \nu \rho \sigma} (B^\dagger_\nu \partial_\rho B_\sigma - B_\nu \partial_\rho B^\dagger_\sigma) \bar q \gamma_\mu \gamma^5 q$ &
  0 & $v^{\bot 2}$ (SM); $q^2$ (DM) & Yes \\
\hline \hline
\end{tabular}
\caption{The momentum ($q$) or velocity ($v^\bot$) suppression of spin-independent or spin-dependent scattering cross sections mediated by
operators V$(7-10)_-$.  Each suppression is labeled based on whether it arises from the Standard Model (SM) or dark matter (DM) bilinear.
Also indicated is if the operator permits $s$-wave annihilation.  If a cross section contains multiple terms with different kinematic
suppressions, then they are listed on separate lines.}
\label{table:PrimedOpsMaster}
\normalsize
\end{center}
\end{table}




\end{document}